\documentclass{JINST}

\usepackage{longtable}
\usepackage{multirow} 
\usepackage{lscape}

\usepackage{lineno}

\title{Radiopurity assessment of the tracking readout for the NEXT double beta decay experiment}

\author{S.~Cebri\'an,$^{a,b}$\thanks{Corresponding author
(scebrian@unizar.es).} J.~P\'erez,$^{c}$ I.~Bandac,$^{b}$
L.~Labarga,$^{d}$ V.~\'Alvarez,$^{e}$ A.I.~Barrado,$^{f}$
A.~Bettini,$^{b,g}$ F.I.G.M.~Borges,$^{h}$ M.~Camargo,$^{i}$
S.~C\'arcel,$^{e}$
 A.~Cervera,$^{e}$ C.A.N.~Conde,$^{h}$
E.~Conde,$^{f}$ T.~Dafni,$^{a,b}$ J.~D\'iaz,$^{e}$ R.~Esteve,$^{j}$
L.M.P.~Fernandes,$^{h}$ M.~Fern\'andez,$^{f}$ P.~Ferrario,$^{e}$
A.L.~Ferreira,$^{k}$ E.D.C.~Freitas,$^{h}$ V.M.~Gehman,$^{l}$
A.~Goldschmidt,$^{l}$
J.J.~G\'omez-Cadenas,$^{e}$\thanks{Spokesperson
(gomez@mail.cern.ch).} D.~Gonz\'alez-D\'iaz,$^{a,b}$
R.M.~Guti\'errez,$^{i}$ J.~Hauptman,$^{m}$ J.A.~Hernando
Morata,$^{n}$ D.C.~Herrera,$^{a,b}$ I.G.~Irastorza,$^{a,b}$
A.~Laing,$^{e}$ I.~Liubarsky,$^{e}$ N.~L\'opez-March,$^{e}$
D.~Lorca,$^{e}$ M.~Losada,$^{i}$ G.~Luz\'on,$^{a,b}$
A.~Mar\'i,$^{j}$ J.~Mart\'in-Albo$^{e}$ A.~Mart\'inez,$^{e}$
G.~Mart\'inez-Lema,$^{n}$ T.~Miller,$^{l}$ F.~Monrabal,$^{e}$
M.~Monserrate,$^{e}$ C.M.B.~Monteiro,$^{h}$ F.J.~Mora,$^{j}$ L.M.
Moutinho,$^{k}$ J.~Mu\~noz Vidal,$^{e}$ M.~Nebot-Guinot,$^{e}$
D.~Nygren,$^{l}$ C.A.B.~Oliveira,$^{l}$ A. Ortiz de
Sol\'orzano,$^{a,b}$ J.L.~P\'erez Aparicio,$^{o}$ M.~Querol,$^{e}$
J.~Renner,$^{l}$ L.~Ripoll,$^{p}$ J.~Rodr\'iguez,$^{e}$
F.P.~Santos,$^{h}$ J.M.F.~dos Santos,$^{h}$ L.~Serra,$^{e}$
D.~Shuman,$^{l}$ A. Sim\'on,$^{e}$ C.~Sofka,$^{q}$ M.~Sorel,$^{e}$
J.F.~Toledo,$^{j}$ J.~Torrent,$^{p}$ Z.~Tsamalaidze,$^{r}$
J.F.C.A.~Veloso,$^{k}$ J.A.~Villar,$^{a,b}$ R.C.~Webb,$^{q}$
J.T.~White,$^{q}$ N.~Yahlali$^{e}$
\\
\llap{$^{a}$}
Laboratorio de F\'isica Nuclear y Astropart\'iculas, Universidad de Zaragoza\\
Calle Pedro Cerbuna 12, 50009 Zaragoza, Spain\\
\llap{$^b$}
Laboratorio Subterráneo de Canfranc\\
Paseo de los Ayerbe s/n, 22880 Canfranc Estación, Huesca, Spain\\
\llap{$^{c}$}
Instituto de F\'isica Te\'orica (IFT), UAM/CSIC\\
Campus de Cantoblanco, 28049 Madrid, Spain\\
\llap{$^{d}$}
Departamento de F\'isica Te\'orica, Universidad Aut\'onoma de Madrid\\
Campus de Cantoblanco, 28049 Madrid, Spain\\
\llap{$^{e}$}
Instituto de F\'isica Corpuscular (IFIC), CSIC \& Universitat de Val\`encia\\
Calle Catedr\'atico Jos\'e Beltr\'an, 2, 46980 Paterna, Valencia, Spain\\
\llap{$^{f}$}
Centro de Investigaciones Energ\'eticas, Medioambientales y Tecnol\'ogicas (CIEMAT)\\
Complutense 40, 28040 Madrid, Spain \\
\llap{$^{g}$} Padua University and INFN Section, Dipartimento di
Fisca G. Galilei, Via Marzolo 8, 35131 Padova, Italy\\
\llap{$^{h}$}
Departamento de Fisica, Universidade de Coimbra\\
Rua Larga, 3004-516 Coimbra, Portugal\\
\llap{$^{i}$}
Centro de Investigaciones en Ciencias B\'asicas y Aplicadas, Universidad Antonio Nari\~no\\
Carretera 3 este No.\ 47A-15, Bogot\'a, Colombia\\
\llap{$^{j}$}
Instituto de Instrumentaci\'on para Imagen Molecular (I3M), Universitat Polit\`ecnica de Val\`encia\\
Camino de Vera, s/n, Edificio 8B, 46022 Valencia, Spain\\
\llap{$^{k}$}
Institute of Nanostructures, Nanomodelling and Nanofabrication (i3N), Universidade de Aveiro\\
Campus de Santiago, 3810-193 Aveiro, Portugal\\
\llap{$^{l}$}
Lawrence Berkeley National Laboratory (LBNL)\\
1 Cyclotron Road, Berkeley, California 94720, USA\\
\llap{$^{m}$}
Department of Physics and Astronomy, Iowa State University\\
12 Physics Hall, Ames, Iowa 50011-3160, USA\\
\llap{$^{n}$}
Instituto Gallego de F\'isica de Altas Energ\'ias (IGFAE), Univ.\ de Santiago de Compostela\\
Campus sur, R\'ua Xos\'e Mar\'ia Su\'arez N\'u\~nez, s/n, 15782 Santiago de Compostela, Spain\\
\llap{$^{o}$}
Dpto.\ de Mec\'anica de Medios Continuos y Teor\'ia de Estructuras, Univ.\ Polit\`ecnica de Val\`encia\\
Camino de Vera, s/n, 46071 Valencia, Spain\\
\llap{$^{p}$}
Escola Polit\`ecnica Superior, Universitat de Girona\\
Av.~Montilivi, s/n, 17071 Girona, Spain\\
\llap{$^{q}$}
Department of Physics and Astronomy, Texas A\&M University\\
College Station, Texas 77843-4242, USA\\
\llap{$^{r}$}
Joint Institute for Nuclear Research (JINR)\\
Joliot-Curie 6, 141980 Dubna, Russia\\
\\
}

\abstract{The ``Neutrino Experiment with a Xenon Time-Projection
Chamber'' (NEXT) is intended to investigate the neutrinoless double
beta decay of $^{136}$Xe, which requires a severe suppression of
potential backgrounds; therefore, an extensive screening and
selection process is underway to control the radiopurity levels of
the materials to be used in the experimental set-up of NEXT. The
detector design combines the measurement of the topological
signature of the event for background discrimination with the energy
resolution optimization. Separate energy and tracking readout planes
are based on different sensors: photomultiplier tubes for
calorimetry and silicon multi-pixel photon counters for tracking.
The design of a radiopure tracking plane, in direct contact with the
gas detector medium, was specially challenging since the needed
components like printed circuit boards, connectors, sensors or
capacitors have typically, according to available information in
databases and in the literature, activities too large for
experiments requiring ultra-low background conditions. Here, the
radiopurity assessment of tracking readout components based on
gamma-ray spectroscopy using ultra-low background germanium
detectors at the Laboratorio Subterráneo de Canfranc (Spain) is
described. According to the obtained results, radiopure enough
printed circuit boards made of kapton and copper, silicon
photomultipliers and other required components, fulfilling the
requirement of an overall background level in the region of interest
of at most $8\times10^{-4}$ counts keV$^{-1}$ kg$^{-1}$ y$^{-1}$,
have been identified.}

\keywords{Double beta decay; Time-Projection Chamber (TPC); Gamma
detectors (HPGe); Search for radioactive material}

\begin{document}
\section{Introduction}

\begin{figure}
\begin{center}
  \includegraphics[height=.3\textheight]{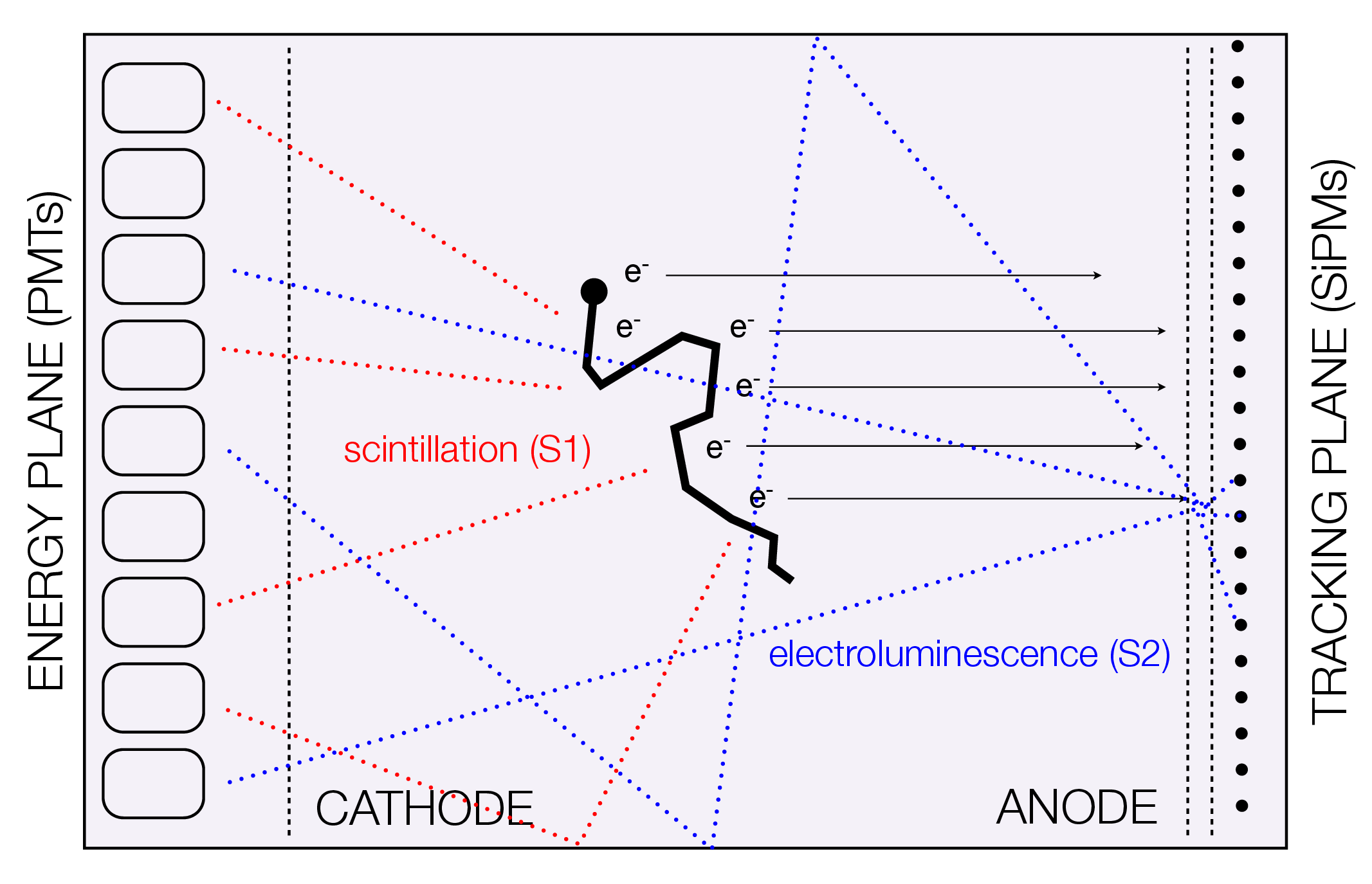}
  \caption{Concept of the NEXT experiment: light from the Xe electroluminescence generated
at the anode is recorded both in the photosensor plane right behind
it for tracking and in the photosensor plane behind the transparent
cathode for a precise energy measurement. Primary scintillation
defining the start of the event is also detected by the cathode
photosensors.}
  \label{soft}
\end{center}
\end{figure}

Double beta decay experiments are one of the most active research
topics in Neutrino Physics. The observation of the neutrinoless
mode, as a peak at the transition energy, could give unique
information on the neutrino nature, showing that neutrinos are
Majorana particles, and for the determination of their mass
hierarchy (see for instance \cite{dbdrefs1}-\cite{dbdrefs3}). The
current generation of experiments aims at detector target masses at
the 100 kg scale, while the next generation will need to go to the
ton scale in order to completely explore the inverse hierarchy
models of neutrino mass \cite{dbdexp}. Since double beta decay is a
very rare process, an ultra-low background level at the region where
the signal is expected to appear is one of the main experimental
requirements for a successful experiment. The NEXT experiment
(``\underline{N}eutrino \underline{E}xperiment with a
\underline{X}enon \underline{T}ime-Projection Chamber'') aims to
search for such a decay in $^{136}$Xe at the Laboratorio Subterráneo
de Canfranc (LSC) \cite{lsc}, located at the Spanish Pyrenees, with
a source mass of $\sim$100 kg (NEXT-100 phase). NEXT takes a
detector$=$source approach, with the double beta emitters inside the
detector, in order to maximize signal detection efficiency and the
accumulation of a large mass of the relevant isotope. The challenge
of NEXT is to combine the measurement of the topological signature
of the event (in order to discriminate the signal from background)
with the energy resolution optimization (to single out the peak at
the sum energy of the two emitted electrons). The NEXT detector will
be a high pressure gaseous xenon Time-Projection Chamber (TPC) with
proportional electroluminescent (EL) amplification \cite{next}:
light from the Xe electroluminescence generated at the anode is
recorded both in the photosensor plane right behind it for tracking
and in the photosensor plane behind the transparent cathode for a
precise energy measurement. As illustrated in figure~\ref{soft}, the
separate energy and tracking readout planes, located at opposite
sides of the pressure vessel, will use different sensors:
photomultiplier tubes (PMTs) for calorimetry (and for fixing the
start of the event) and silicon photomultipliers (SiPMs) for
tracking. While work on prototypes is still ongoing
\cite{berkeley}-\cite{protomm2}, the installation of shielding and
ancillary system started at LSC in 2013. Underground commissioning
of the NEW detector begun at the end of 2014 and first data are expected along 2015. The NEW (NEXT-WHITE)
apparatus\footnote{The name honours the memory of the late Professor
James White, key scientist of the NEXT project.} is the first phase
of the NEXT detector to operate underground; it is a downscale 1:2
in size (1:8 in mass) of NEXT-100.

The goal of NEXT is to explore electron neutrino effective Majorana
masses below 100 meV for a total exposure of 500 kg$\cdot$year
\cite{dbdrefs1}. To reach this sensitivity, there are two basic
requirements \cite{sense}: 1) An energy resolution of at most 1\%
FWHM at the transition energy (Q$_{\beta\beta}=$2.458 MeV), which is
reachable with EL amplification according to the results of
prototypes \cite{berkeley,lorca}. 2) A background level below
$8\times10^{-4}$ counts keV$^{-1}$ kg$^{-1}$ y$^{-1}$ in the energy
region of interest, achievable thanks to passive shieldings,
background discrimination techniques based on charged particle
tracking and a thorough material radiopurity control. The NEXT-100
shielding will consist of a 20-cm-thick lead castle covering the
pressure vessel, suppressing by more than four orders of magnitude
the external gamma radiation at the region of interest, together
with an additional 12-cm-thick layer of copper inside the vessel
intended to shield emissions from the vessel itself. The ability to
discriminate signal from background is a powerful tool in NEXT.
Signal events will appear uniformly distributed in the source volume
of enriched xenon and will have a distinctive topology (a twisted
long track, about 30 cm long at 10 bar, ending in two larger energy
depositions). Defining a fiducial volume eliminates all charged
backgrounds entering the detector while confined tracks generated by
neutral particles, like high-energy gammas, can be suppressed by
reconstructed topology of the events. Thus, the relevance of a
background source depends on its probability of generating a
signal-like track in the fiducial volume with energy around
Q$_{\beta\beta}$. The most dangerous background sources are then
$^{208}$Tl and $^{214}$Bi, isotopes of the progeny of $^{232}$Th and
$^{238}$U.

Concerning the radiopurity control, an extensive material screening
and selection process for NEXT components is underway for several
years. Determination of the activity levels is based on gamma-ray
spectroscopy using ultra-low background germanium detectors at LSC
and also on other techniques like Glow Discharge Mass Spectrometry
and Inductively Coupled Plasma Mass Spectrometry. Materials to be
used in the shielding, pressure vessel, electroluminescence and high
voltage components, and energy and tracking readout planes have been
taken into consideration and first results have been presented in
\cite{jinstrp,aiprp,icheprp}. These results are the input for the
construction of a precise background model of the NEXT experiment
based on Monte Carlo simulations, which is currently underway using the
Geant4 code \cite{ichep}; preliminary estimates indicate that
shielding, vessel, field cage and energy plane could produce a
background level of $\sim4\times10^{-4}$ counts keV$^{-1}$ kg$^{-1}$
y$^{-1}$ in the region of interest. The design of a radiopure
tracking readout plane is complicated by the fact that printed
circuit boards and electronic components, involving typically different composite materials, show in many cases activity levels too large for
being used in experiments demanding ultra-low background conditions
(see for instance \cite{heusser} or \cite{ILI11}). SiPM technology offers an outstanding performance for photon detection \cite{SiPMstech}, but scarce information on its radiopurity is available.
According to simulations, a maximum activity of 70 mBq from $^{208}$Tl and
$^{214}$Bi at the tracking plane could be tolerated, since it would
generate other $4\times10^{-4}$ counts keV$^{-1}$ kg$^{-1}$ y$^{-1}$
in the region of interest. Therefore, an exhaustive screening
program specifically for the tracking readout components was undertaken and
is described here.

The tracking function in NEXT-100 detector will be provided by a
plane of SiPMs operating as sensor pixels and placed behind the
transparent EL gap (see figure~\ref{soft}), inside the pressure
vessel. The SiPMs will be mounted in an a array of 107 square boards
(named ``Dice Boards'', DB), to cover the whole field cage cross
section. Each DB contains 8$\times$8 SiPM sensors with a pitch of
$\sim$1~cm between them; in addition, one NTC thermistor acting as
temperature sensor is placed on the center of each DB and LEDs are
included to allow a precise PMT geometrical calibration. As SiPMs
provide very small current signals, the transmission of a high
number of these signals from the photodetectors to the front-end
electronics, crossing through the pressure vessel and traveling
along several meters of cables, is not easy. The front-end
electronics should be placed as close as possible to the detector;
it will be located outside the lead shielding to minimize
backgrounds from their non-radiopure components. The design of the
tracking readout plane and front-end electronics has been tested in
NEXT-DEMO \cite{protodemo,sipms}.

The structure of the paper is the following. Section~\ref{meas}
summarizes all the measurements performed, describing both the
samples analyzed and the detectors used. Activity results obtained
are collected in section~\ref{resu}, together with the discussion of implications for design and for the NEXT-100 background model. Finally, conclusions are drawn in section~\ref{disc}.


\section{Measurements}
\label{meas}

The material screening program of the tracking readout of the NEXT experiment is
based on germanium $\gamma$-ray spectrometry using ultra-low
background detectors operated deep underground, at a depth of 2450
m.w.e., from the Radiopurity Service of LSC; being a non-destructive
technique, the actual components to be used in the experiment can be
analyzed.

The Radiopurity Service of LSC offers several detectors to measure
ultra-low level radioactivity. They are p-type close-end coaxial
2.2-kg High Purity germanium detectors, from Canberra France, with
aluminum or copper cryostats and 100-110\% relative
efficiencies\footnote{Efficiency relative to a $3''\times3''$ NaI
detector at 1332 keV and for a distance of 25 cm between source and
detector.}. Data acquisition is based on Canberra DSA 1000 modules
and shielding consists of 5 or 10 cm of copper in the inner part
surrounded by 20 cm of low activity lead, flushed with nitrogen gas to
avoid airborne radon intrusion. The measurements related with the
tracking readout were carried out at LSC using in particular four
different $\sim$2.2 kg detectors from LSC (named GeAnayet, GeAspe,
GeLatuca, and GeOroel) and also a $\sim$1 kg detector from the
University of Zaragoza (named Paquito). For the measurements
presented here, only GeAspe had a 10-cm-thick copper shield.
Table~\ref{gerates} shows the counting rates of all the detectors
used in the energy window from 100 to 2700 keV and at different
peaks: 583 keV from $^{208}$Tl, 609 keV from $^{214}$Bi and 1461 keV
from $^{40}$K; all rates are expressed in counts per day and per kg
of germanium detector. More details on detectors and their
backgrounds can be found at \cite{jinstrp,paquito}.

\begin{table}
\caption{Background counting rates (expressed in counts d$^{-1}$
kg$^{-1}$) of the germanium detectors used at LSC for the NEXT
tracking plane measurements. Integral rate from 100 to 2700 keV and
rates at different peaks (583 keV from $^{208}$Tl, 609 keV from
$^{214}$Bi and 1461 keV from $^{40}$K) are presented. Only
statistical errors are quoted.}
\begin{tabular}{lcr@{$\pm$}lccc}
\\
\hline

Detector name  & Mass (kg) &  \multicolumn{2}{c}{100-2700 keV} & 583
keV &609 keV &1461 keV  \\ \hline

GeAnayet &     2.183 &  714&3 &  3.73$\pm$0.40 &   1.76$\pm$0.28& 0.31$\pm$0.20 \\

GeAspe & 2.187 & 441&2  & 3.77$\pm$0.47 & 3.74$\pm$0.45 & 0.58$\pm$0.24 \\


GeLatuca  &  2.187 &  667&3  & 3.02$\pm$0.32 & 5.66$\pm$0.39 & 0.47$\pm$0.13 \\

GeOroel & 2.230 &   461&2 &    0.98$\pm$0.23 &   2.69$\pm$0.30 & 0.32$\pm$0.13\\

 Paquito  & 1 &  79&2 &  0.27$\pm$0.09 & 0.48$\pm$0.21 & 0.25$\pm$0.13 \\

 \hline
\end{tabular}

\label{gerates}
\end{table}

\begin{figure}
\begin{center}
  \includegraphics[height=.3\textheight]{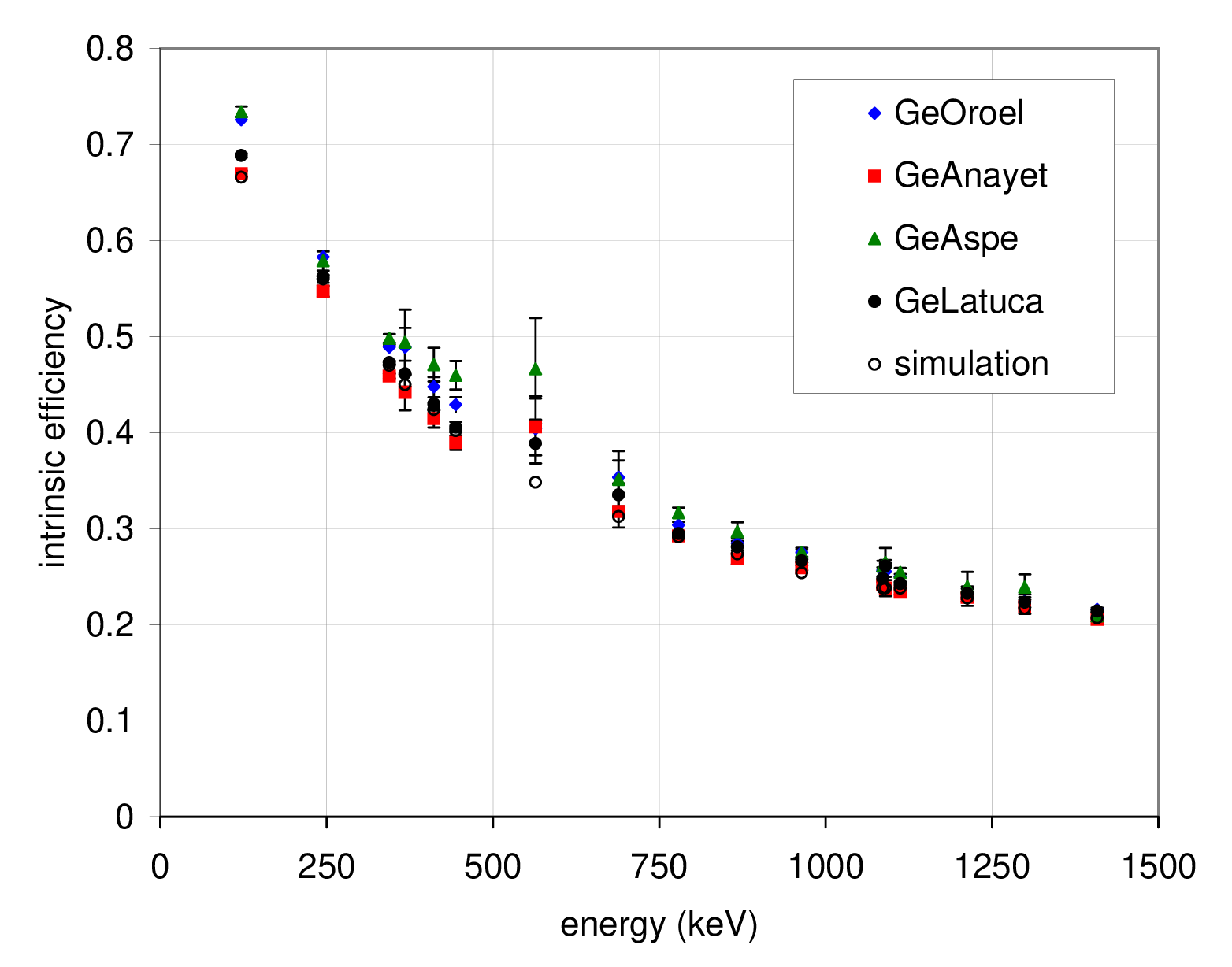}
  \caption{Intrinsic efficiency measured with a $^{152}$Eu reference point source for the 2-kg germanium detectors of the Radiopurity Service of LSC used in the measurements and the corresponding simulation (for Anayet detector).}
  \label{efcurves}
\end{center}
\end{figure}

To derive the activity of an isotope producing a gamma emission of a
certain energy in a sample, the main ingredients are the net signal
(that is, the number of events at the gamma line stemming from the
sample) and the full-energy peak detection efficiency at the
corresponding energy. The criteria proposed in Currie's landmark
paper \cite{currie} and revised in \cite{gator,revisiting} have been
followed to evaluate net signals; activities have been quantified
when possible and upper limits with a 95.45\%~C.L. have been derived
otherwise. In the cases when the background from the experimental
setup gives the dominant contribution to the gamma-line under
evaluation, the gaussian limit has been taken in the statistical
analysis of that line. Concerning the estimate of the detection
efficiency, Monte Carlo simulations based on the Geant4
\cite{geant4} code have been performed for each sample, accounting
for intrinsic efficiency, the geometric factor and self-absorption
at the sample. The configuration of each detector, including the
germanium crystal, cryostat and components inside, is implemented in
detail in the code following the manufacturer's specifications; the
external copper and lead shielding is also considered.
G4EmLivermorePhysics class is used to define the physics models for
the simulation of the gamma emissions. Validation of the simulation
environment has been made by comparing the efficiency curve of each
detector (measured with a $^{152}$Eu reference point source of known
activity located at a distance of 25 cm) with the simulated one.
Figure~\ref{efcurves} shows the intrinsic efficiency (corrected by
solid angle) obtained for GeOroel, GeTobazo, GeAnayet and GeLatuca
detectors together with a simulation considering GeAnayet geometry.
The inclusion in the simulations of a dead layer over the whole
crystal, with a thickness from 0.5 to 1 mm according to the detector
specification sheets, has a non-negligible effect on the detector
efficiency and allows to improve the agreement with measurements,
especially at low energies, reducing deviations to a level of 5\%.
The relative efficiencies derived from measurements and simulation
reproduce the values specified by the manufacturer. The overall
uncertainty in the calculated detection efficiency of the samples
for every gamma line is estimated to be 10\%. The activity values
and limits presented in table \ref{rpm} are further affected by this
uncertainty.

Activities of different sub-series in the natural chains of
$^{238}$U, $^{232}$Th and $^{235}$U as well as of common primordial,
cosmogenic or anthropogenic radionuclides like $^{40}$K, $^{60}$Co
and $^{137}$Cs have been evaluated by analyzing the most intense
gamma lines of different isotopes. For $^{238}$U, emissions from
$^{234}$Th and $^{234m}$Pa are searched to quantify activity of the
upper part of the chain\footnote{The low intensity of these emissions makes upper limits on activity obtained from their analysis be typically much higher than for other isotopes in the chain.} and lines from $^{214}$Pb and $^{214}$Bi for
the sub-chain starting with $^{226}$Ra up to $^{210}$Pb. For
$^{232}$Th chain, emissions of $^{228}$Ac are analyzed for the upper
part and those of $^{212}$Pb, $^{212}$Bi and $^{208}$Tl for the
lower one. Concerning $^{235}$U chain, only emissions from the
parent isotope are taken into account; even in the cases of not
detailed quantification, activities from $^{235}$U are always found
reasonably consistent with natural abundance.

Table~\ref{gedata} summarizes the measurements performed for the
samples analyzed in this work, indicating material and supplier, the
detector used, the size of the sample and the live time of data taking.
Most of the samples were cleaned in an ultrasonic bath and with pure
alcohol before starting the screening.

\begin{table}
\caption{Information on measurements performed: component and
supplier, detector used, samples size (mass or number of pieces) and
screening live time. The corresponding row number of table~3 where
the activity values obtained for each sample are reported is also
quoted.}
\begin{tabular}{lccr@{ }lr@{.}l}
\\
\hline
Component, Supplier & \# in table~3 &  Detector & \multicolumn{2}{c}{Sample size} &  \multicolumn{2}{c}{Time (d)} \\
\hline


Cuflon, Polyflon & 1 & GeOroel & 1876&g & 24&29 \\
Bonding film, Polyflon & 2 & GeAnayet & 288&g & 30&83 \\
Cuflon Dice Board, Pyrecap & 3 & GeOroel & 140&g & 45&11 \\
Kapton-Cu Dice Board, Flexiblecircuits & 4 & GeOroel& 647&g & 26&60\\
\hline

FFC/FCP connector, Hirose & 5 & Paquito & 19 pc $\times$&1.23 g/pc&  6&83\\
P5K connector, Panasonic  & 6 & Paquito & 15 pc $\times$&0.67 g/pc & 7&58 \\
Thermoplastic connector, Molex & 7 & GeLatuca & 29 pc $\times$&0.53
g/pc & 17&20 \\ \hline

Solder paste, Multicore & 8 & GeLatuca & 457&g & 44&30 \\
Solder wire, Multicore & 9 & Paquito & 91&g & 7&74 \\
Silver epoxy, Circuit Works & 10 & GeLatuca & 125&g & 55&11 \\
\hline

SiPMs 1$\times$1 mm$^{2}$, SensL & 11 & GeAspe & 102 pc $\times$&3.4 mg/pc & 41&42 \\
SiPMs 6$\times$6 mm$^{2}$, SensL & 12 & GeAspe & 99 pc $\times$&96 mg/pc & 59&62 \\
\hline

NTC sensor, Murata & 13 & GeLatuca & 1000 pc $\times$&4.5 mg/pc & 28&27\\
LEDs, Osram & 14 & GeLatuca & 989 pc $\times$&0.8 mg/pc & 32&35\\
Plexiglas/PMMA, Evonik & 15 & GeLatuca & 1669&g & 48&87\\
Ta capacitor, Vishay Sprague  & 16 & GeAnayet & 277 pc $\times$&0.64 g/pc & 17&97 \\

 \hline
\end{tabular}
 \label{gedata}
\end{table}

\section{Results}
\label{resu}

\footnotesize
\begin{landscape}
\label{rpm}
\begin{longtable}{p{0.2cm}p{2.9cm}p{2.0cm}p{1.0cm}p{1.3cm}p{1.5cm}p{1.5cm}p{1.5cm}p{1.2cm}p{1.2cm}p{1cm}p{1cm}}

\hline
\textbf{\#} & \textbf{Component} & \textbf{Supplier}  & \textbf{Unit} & \textbf{$^{238}$U} & \textbf{$^{226}$Ra} & \textbf{$^{232}$Th} &\textbf{$^{228}$Th} & \textbf{$^{235}$U}& \textbf{$^{40}$K}  & \textbf{$^{60}$Co}& \textbf{$^{137}$Cs}\\
 \hline
\endfirsthead
(Continuation)\\
\hline
\textbf{\#} & \textbf{Component} & \textbf{Supplier} & \textbf{Unit} & \textbf{$^{238}$U} & \textbf{$^{226}$Ra} &\textbf{$^{232}$Th} &\textbf{$^{228}$Th} & \textbf{$^{235}$U}& \textbf{$^{40}$K}  & \textbf{$^{60}$Co}& \textbf{$^{137}$Cs}\\
\hline
\endhead
&(Follows at next page)\\
\endfoot
\endlastfoot

1 & Cuflon & Polyflon &  mBq/kg & $<$33 & $<$1.3 & $<$1.1 & $<$1.1  & $<$0.6 & 4.8$\pm$1.1 & $<$0.3 & $<$0.3  \\
2 & Bonding film & Polyflon &  mBq/kg & 1140$\pm$300 &487$\pm$23 & 79.8$\pm$6.6 & 66.0$\pm$4.8 &  & 832 $\pm$87 & $<$4.4 & $<$3.8 \\
3 & Cuflon Dice Board & Pyrecap & mBq/pc & $<$ 7.6 & 0.28$\pm$0.08 & $<$ 0.28 &  $<$ 0.16 & $<$0.13 & $<$1.2   & $<$0.07 & $<$ 0.06  \\
4 & Kapton-Cu Dice Board &Flexiblecircuits&  mBq/pc&   $<$1.3&0.031$\pm$0.004    &0.027$\pm$0.008&   0.042$\pm$0.004 && 12.1$\pm$1.2&   $<$0.01&   $<$0.01\\
\hline

5 & FFC/FCP connector & Hirose &  mBq/pc & $<$50 & 4.6$\pm$0.7 & 6.5$\pm$1.2 &6.4$\pm$1.0 & $<$0.75 & 3.9$\pm$1.4 & $<$0.2 & $<$0.5  \\
6 & P5K connector & Panasonic &  mBq/pc & $<$42 & 6.0$\pm$0.9 & 9.5$\pm$1.7 & 9.4$\pm$1.4 & $<$0.95 & 4.1$\pm$1.5 & $<$0.2 & $<$0.8\\
7 & Thermopl. connector & Molex &  mBq/pc & $<$7.3 & 1.77$\pm$0.08 & 3.01$\pm$0.19 & 2.82$\pm$0.15 & $<$0.31 & 2.12$\pm$0.25 & $<$0.022 & 0.27$\pm$0.03\\
\hline

8 & Solder paste & Multicore &  mBq/kg & $<$310 & $<$2.7 & $<$4.7 & $<$2.5 & $<$5.2 & $<$13 & $<$1.0 & $<$1.6 \\
9 & Solder wire & Multicore &  mBq/kg & $<$4900 & (7.7$\pm$1.2)10$^{2}$ & $<$147 & $<$14 & & $<$257 & $<$30 & $<$36 \\
10& Silver epoxy&   Circuit Works&  mBq/kg& $<$1.0 10$^{3}$&   13.6$\pm$2.8&   $<$18& $<$ 16& $<$4.5&    $<$52& $<$1.9&    $<$2.2\\
\hline

11& SiPMs 1$\times$1 mm$^{2}$ & SensL & $\mu$Bq/pc & $<$320 & $<$2.7
& $<$6.9 & $<$2.0 &$<$1.0 & $<$16 & $<$0.8 &$<$2.0 \\
12& SiPMs 6$\times$6 mm$^{2}$ & SensL & $\mu$Bq/pc & $<$410 & $<$3.2
& $<$12 & $<$2.8 &$<$2.5 & $<$25 & $<$1.2 &$<$1.3 \\ \hline

13& NTC sensor&    Murata  &    $\mu$Bq/pc&   $<$96& $<$0.8&    $<$0.9 &$<$0.3&   $<$0.3&    $<$2.9&    $<$0.2&    $<$0.2\\
14& LED&   Osram   &    $\mu$Bq/pc&   $<$90  &1.4$\pm$0.2    &3.5$\pm$0.4    &3.0$\pm$0.3&   $<$0.6&    $<$4.0&    $<$0.2 &$<$0.3\\
15& Plexiglas/PMMA&   Evonik&  mBq/kg& $<$208 &$<$1.3&   $<$2.2 &$<$1.0&   $<$1.1 &$<$8.1&   $<$0.4 &$<$0.6\\
16& Ta capacitor & Vishay Sprague & mBq/pc & $<$0.8 & 0.043$\pm$0.003 & 0.034$\pm$0.004 & 0.032$\pm$0.003 & $<$ 0.010 & & $<$0.002  & $<$0.003 \\

\hline

\caption{Activities measured for tracking readout components to be
used in NEXT. Results reported for $^{238}$U and $^{232}$Th
correspond to the upper part of the chains and those of $^{226}$Ra
and $^{228}$Th give activities of the lower parts.  All the
activities have a global 10\% uncertainty coming from the Monte
Carlo estimate of the detection efficiency (see text for more
details).}
\end{longtable}
\end{landscape}
\normalsize

The activity results obtained for the samples analyzed dealing with
the tracking readout plane are all summarized in table~\ref{rpm};
reported errors correspond to $1\sigma$ uncertainties including both
statistical and efficiency uncertainties. The expected contribution
to the background level in the region of interest of NEXT-100 from
the activities of the most relevant components of the tracking plane
has been evaluated by Monte Carlo simulation and is reported in
table~\ref{comcont}. In the following, each sample is described and
the corresponding results discussed.

\begin{table}
\caption{Summary of activities measured for the most relevant
components of the tracking readout plane and the corresponding
expected background rate in the region of interest for NEXT-100
estimated by Monte Carlo simulation. As a reference, the maximum
allowed activities and rate for the whole tracking plane are
indicated in the first row. Last column indicates if the component
has been finally accepted or not for use in NEXT-100 set-up.}
\begin{tabular}{lcccc}
\\
\hline
Component& $^{208}$Tl activity & $^{214}$Bi activity & Background rate b & Accepted \\
& & & (counts keV$^{-1}$ kg$^{-1}$ y$^{-1}$) & \\ \hline

Whole plane & 35 mBq & 35 mBq & 4$\times 10^{-4}$ & \\ \hline

Cuflon Dice Boards & $<$0.06 mBq/pc & 0.28 mBq/pc & (1.8$<b<$2.1$)\times10^{-4}$ & No \\

Hirose connectors & 2.3 mBq/pc & 4.6 mBq/pc & 4.3$\times10^{-3}$ & No \\

Kapton Dice Boards & 0.015 mBq/pc & 0.031 mBq/pc & 2.8$\times10^{-5}$ & Yes \\

SiPMs (1$\times$1 mm$^{2}$) & $<$0.03 $\mu$Bq/pc & $<$0.09 $\mu$Bq/pc & $<0.5\times10^{-5}$ & Yes \\

LEDs & 1.1 $\mu$Bq/pc & 1.4 $\mu$Bq/pc & 0.2$\times10^{-5}$ & Yes \\

NTC sensors & $<$0.1 $\mu$Bq/pc & $<$0.8 $\mu$Bq/pc &
$<0.06\times10^{-5}$ & Yes \\ \hline

\end{tabular}
\label{comcont}
\end{table}

\subsection{Printed Circuit Boards and cables}

Printed Circuit Boards (PCBs) are commonly made of different
materials and a large number of radiopurity measurements can be
found in \cite{ILI11}. Therefore, several options have been taken
into consideration for the substrate of SiPMs arrays. FR4 was
disregarded because of both an unacceptable~high rate of outgassing
and bad radiopurity; glass fiber-reinforced materials at base plates
of circuit boards are generally recognized as a source of
radioactive contamination \cite{heusser}.

Cuflon$\circledR$ offers low activity levels, as shown in the
measurement of samples from Crane
Polyflon\footnote{http://www.polyflon.com} by GERDA \cite{BUD09} and
at \cite{NIS09}, using both ICPMS and Ge gamma spectroscopy. As
presented in \cite{jinstrp}, a measurement of Polyflon cuflon made
of a 3.18-mm-thick PTFE layer sandwiched by two 35-$\mu$m-thick
copper sheets was made for NEXT and results are shown in row \#1 of
table~\ref{rpm}. Adhesive films to glue cuflon sheets are used to
prepare multilayer PCBs; a sample of bonding films made of a
polyolefin co-polymer and supplied also by Crane Polyflon were
screened and results are presented in row \#2 of table~\ref{rpm}.
Four cuflon DB produced by Pyrecap company using these Polyflon
materials were screened. Each DB, with a surface of 79$\times$79
mm$^{2}$ and a mass of 35 g, was made of three cuflon sheets glued
with two bonding films; results are shown in row \#3 of
table~\ref{rpm}, being fully consistent with the individual
measurements of components. Total activity from each cuflon DB was
too high for NEXT requirements, since they could produce a
background of $2.1\times10^{-4}$ counts keV$^{-1}$ kg$^{-1}$
y$^{-1}$ in the region of interest (see table~\ref{comcont});
consequently, other option was searched for.

\begin{figure}
\begin{center}
  \includegraphics[width=.5\textwidth]{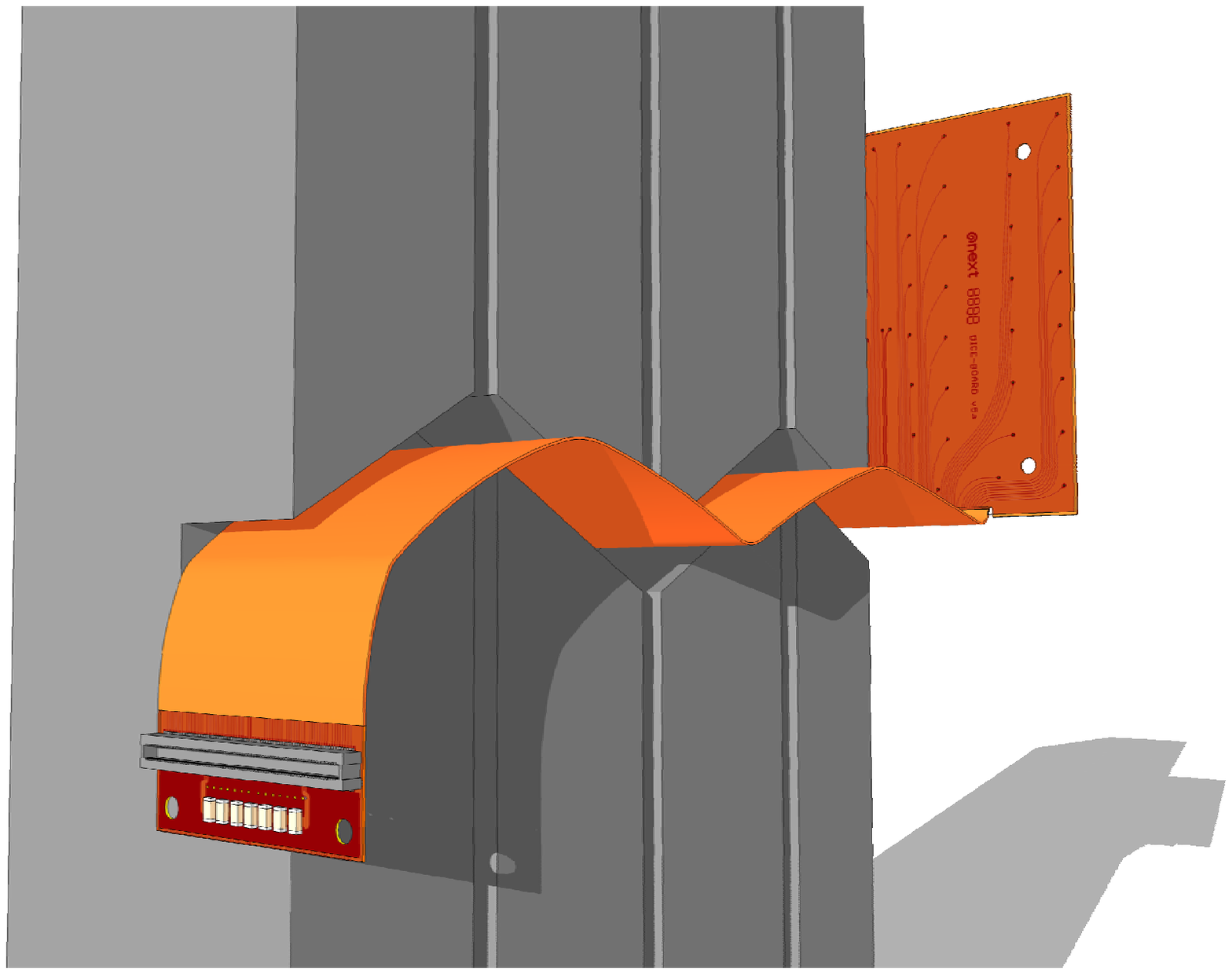} 
  \includegraphics[width=.4\textwidth]{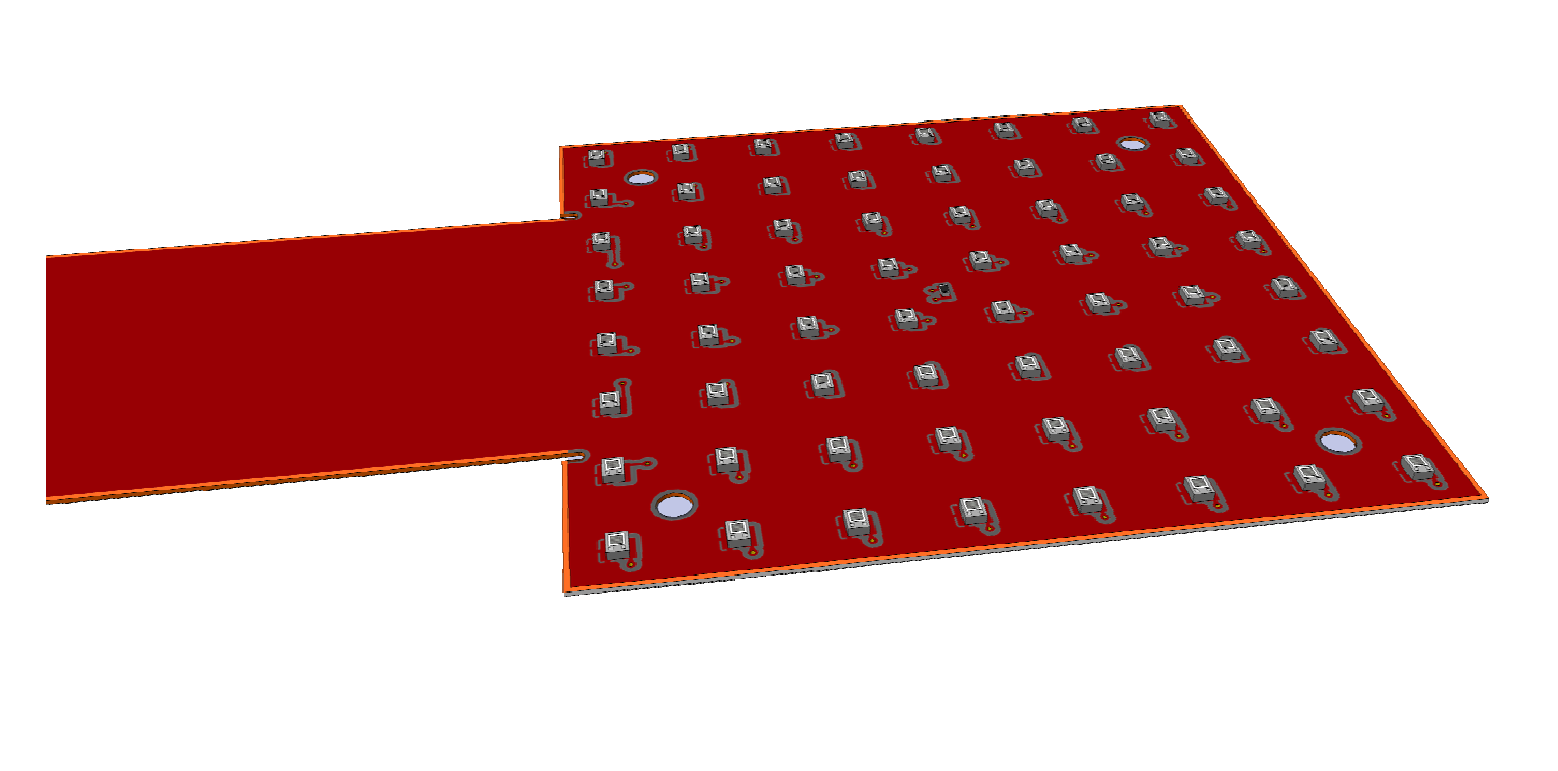}
  \caption{Left: Design of the all-in-one DB made of kapton and copper, consisting of a square part and a long, flexible tail, which allows to locate the connectors behind the inner copper shielding. Right: Detail of the square part where SiPMs and NTC sensor are fixed.}
  \label{kaptondb}
\end{center}
\end{figure}

Components made of just kapton (like cirlex) and copper offer very
good radiopurity, as shown in the measurements of kapton-copper
foils in \cite{radiopuritymm,paquito}. Therefore, new DB produced by
Flexiblecircuit using only kapton, metallized copper and adhesive
were analyzed. A two layer adhesiveless base substrate with
polyimide coverlay on both sides, which only requires a little
amount of adhesive, was chosen for the boards manufacturing. As
shown in figure~\ref{kaptondb}, each DB consists of a square part
with 8 cm side, where SiPMs are fixed, and a long, flexible tail,
which allows to locate connectors behind the inner copper shielding.
The mass of each kapton DB is 16.7 g. A total of 12 units, together
with residual pieces from production to increase the mass sample,
were screened. Results normalized to the DB part actually exposed to
the detector are presented in row \#4 of table~\ref{rpm}. Although a
higher content of $^{40}$K (of relevance for the study of the double
beta decay mode with neutrino emission) has been observed,
activities for the isotopes in the lower parts of $^{238}$U and
$^{232}$Th chains are almost one order of magnitude lower than for
cuflon DB. As shown in table~\ref{comcont}, the quantified activity
of $^{208}$Tl and $^{214}$Bi from DBs gives a rate of
$2.8\times10^{-5}$ counts keV$^{-1}$ kg$^{-1}$ y$^{-1}$ and
consequently kapton DBs have been chosen as the final option for the
tracking readout substrate. The problem of $^{40}$K activity has
been solved in flexible flat cables made also of kapton and copper
by SOMACIS\footnote{http://www.somacis.com} taking care of all the
materials used; from the results of a first screening of a sample of
these cables and the mass of the exposed DB, the upper limit to
$^{40}$K activity would be 1.4 mBq/pc. No contamination was
quantified neither for isotopes in the lower parts of $^{238}$U and
$^{232}$Th chains, which would translate to upper limits at the
level of a few tenths of mBq/pc for exposed DB.

It is worth noting that the portion of cables inside the pressure
vessel, transporting signals from the connectors at the end of the
kapton DB towards the front-end electronics, will be made using the
same materials that the kapton DB.

\subsection{Connectors}
Information on the radiopurity of different types of connectors is
available at \cite{ILI11,BOR,SNO}. Different kinds of board-to-cable
connectors were measured \cite{jinstrp} and results are reported in
rows \#5-7 of table~\ref{rpm}. In particular, FFC/FCP (Flexible
Printed Circuit \& Flexible Flat Cable) connectors supplied by
Hirose\footnote{http://www.hirose.com} and similar P5K connectors
from Panasonic\footnote{http://www.panasonic-electric-works.com}
were considered, finding activities of at least a few mBq/pc for
isotopes in $^{232}$Th and the lower part of $^{238}$U chains and
for $^{40}$K. Thermoplastic connectors 503066-8011 from
Molex\footnote{http://www.molex.com} were also screened, giving
values slightly smaller but of the same order. Since all these
connectors contain Liquid Crystal Polymer (LCP), it seems that the
activity measured is related to this material. As the activity of
connectors would give an unacceptable high rate in the region of
interest (see table~\ref{comcont} for the Hirose connectors), a
direct bonding of the cables to the cuflon DBs was originally
foreseen; however, in the final design using the all-in-one kapton
DBs, connectors are easily placed behind the inner copper shield.

\subsection{Soldering materials}
Different materials intended to be used to solder electronic
components on boards have been analyzed \cite{jinstrp}. A sample of
lead-free SnAgCu solder paste supplied by Multicore (Ref. 698840)
was screened and results are presented in row \#8 of table
\ref{rpm}. $^{108m}$Ag, induced by neutron interactions and having a
half-life of T$_{1/2}=$438 y, has been identified in the paste, with
an activity of (5.26$\pm$0.40)~mBq/kg, while upper limits of a few
mBq/kg have been set for the common radioactive isotopes;
consequently, some tens of grams of the solder paste could be used without concern.
Solder wire with similar composition from Multicore (Ref. 442578)
was also screened (see row \#9 of table~\ref{rpm}), finding in this
case a high activity of the lower part of the $^{238}$U chain. An
activity of $^{210}$Pb of (1.2$\pm$0.4)$\times$10$^{3}$~Bq/kg was
deduced using the bremsstrahlung emission from its daughter nuclide
$^{210}$Bi \cite{nachab}.

A sample of Circuit Works Conductive Epoxy CW2400 mainly made of
silver was measured. It was prepared at LSC just before screening by
mixing epoxy and hardener following specifications. Results are
presented in row \#10 of table~\ref{rpm}. Activity of $^{108m}$Ag
has been measured in this sample too at a level of
(24.6$\pm$1.6)~mBq/kg. Even though the use of this type of silver
epoxies was finally disregarded for electronic boards, it could be
used for the field cage components.

\subsection{SiPMs}
Although silicon is, as germanium, a very radiopure material with
typical intrinsic activities of $^{238}$U and $^{232}$Th at the
level of few $\mu$Bq/kg \cite{heusser}, materials used in the
substrate or package of the chip can be radioactive. Very low
specific activities have been recently obtained by Neutron
Activation Analysis for bare devices from FBK manufacturer
\cite{nEXO}. Two samples of non-functional SiPMs from
SensL\footnote{http://sensl.com}, reported as MLP (Moulded
Lead-frame Package) plastic SMT elements, were screened. One
consisted of 102 units with a surface of 1$\times$1 mm$^{2}$ each,
and the other of 99 6$\times$6 mm$^{2}$ units. Results are shown in
rows \#11-12 of table~\ref{rpm}; activity has not been quantified
for any isotope and upper limits have been derived. Limits per unit
are very similar for both samples, but since the production process
is the same and the proportion of components scales with area,
results from the large 6$\times$6 mm$^{2}$ units allow to set limits
on activities per surface much more stringent. A preliminary
analysis of a sample of 20 units of SiPMs of type TSV (Through
Silicon Via) also from SensL, 3$\times$3 mm$^{2}$ each and made of
different materials, points to a slightly worse radiopurity;
presence of $^{40}$K has been quantified at a level of 1 mBq/pc. As
presented in table~\ref{comcont}, the expected background from SiPMs
of type MLP is $<0.5\times10^{-5}$ counts keV$^{-1}$ kg$^{-1}$
y$^{-1}$. Therefore, these SiPMs will be considered for NEXT.


\subsection{Other components}

NTC thermistors chip type from Murata Manufacturing Co.
Ltd\footnote{http://www.murata.com}, to be used as temperature
sensors at DB, were screened. Each unit is 1.6 mm long and 0.8 mm
wide. As shown in row \#13 of table~\ref{rpm}, upper limits of a few
$\mu$Bq/pc have been set for the common radioisotopes.

Chip LEDs 0603 supplied by Osram\footnote{http://www.osram.com},
with blue emission at 470 nm (LBQ39E) and made with InGaN
technology, were measured. Each unit has a volume of
1.6$\times$0.8$\times$0.3 mm$^{3}$. Results are presented in row
\#14 of table~\ref{rpm}; high specific activities for $^{40}$K,
$^{232}$Th and $^{238}$U chains have been quantified, despite the
very small mass of the sample, which correspond to levels of a few
$\mu$Bq/pc. In principle, the number of LEDs to be used per DB was
between one to four; but it could be reduced to only 10 units for
the whole plane. Following simulations, assuming one LED and one NTC
sensor per DB, the total contribution at the region of interest due
to these components from the lower part of $^{238}$U and $^{232}$Th
chains will be $<3\times10^{-6}$ counts keV$^{-1}$ kg$^{-1}$
y$^{-1}$ (see table~\ref{comcont}).

SiPMs have high photon detection efficiency in the blue region. For
this reason, they need to be coated with a wavelength shifter, to
shift the UV light of the scintillation of xenon to blue, as the
windows of the PMTs at the energy readout plane. TetraPhenyl
Butadiene (TPB) material from Sigma Aldrich has been successfully
used in NEXT prototypes and according to measurements in \cite{SNO},
taking into account the small quantity to be used (about 20 g for
the whole tracking plane), its radiopurity is good enough. Instead
of directly coating the DBs, an envisaged solution was to place
quartz or PolyMethyl Methacrylate (PMMA) thin windows coated with
TPB in front of DBs. A sample made of 134 PMMA sheets
(79$\times$79$\times$1.5 mm$^{3}$ and a mass of 12.46 g each one)
was screened. Material is reported as Plexiglas GS/XT from Evonik
Industries AG\footnote{http://www.evonik.com}. Results are shown in
row \#15 of table~\ref{rpm}, setting upper limits to the analyzed
radioisotopes. Although these results are not bad, the final option
is to use a quartz anode, having this material also an acceptable
radiopurity \cite{LEO08}.

In a first design of the cuflon DBs, capacitors were needed. Ceramic
capacitors were disregarded for being radioactive \cite{ILI11}.
Tantalum capacitors (Vishay Sprague
597D\footnote{http://www.vishay.com}) were screened at LSC and
results are presented in row \#16 of table \ref{rpm}; activity
levels are lower than for other tantalum capacitors \cite{ILI11}. In
addition to activities shown in table \ref{rpm},
the presence of $^{182}$Ta (beta emitter with Q$=$1814.3 keV and
T$_{1/2}=($114.74$\pm$0.12) days, produced by neutron activation on
$^{181}$Ta) was identified. In any case, in the final design of
kapton-Cu DBs no capacitor is used.

\section{Conclusion}
\label{disc}

A thorough control of material radiopurity is being performed in the
construction of the NEXT double beta decay experiment to be operated
at LSC, mainly based on activity measurements using ultra-low
background gamma-ray spectrometry with germanium detectors of the
Radiopurity Service of LSC. Radiopurity information is helpful not
only for the selection of radiopure enough materials, but also for
the development of the detector background model in combination with
Monte Carlo simulations.

The design of a radiopure tracking readout plane for the NEXT
detection system, which must be in direct contact with the gas
detector medium, is a challenge, since PCB materials and electronic
components can typically have much higher activity levels than those
tolerated in ultra-low background experiments. Selection of
in-vessel components for the tracking plane has been performed in
parallel to its design. SiPMs with low enough activity have been
identified. Regarding the substrate for SiPMs, printed circuit
boards made of kapton and copper have been chosen for their better
radiopurity for $^{238}$U and $^{232}$Th chains in comparison with
cuflon boards; even the reduction of the high content in $^{40}$K
measured seems possible for kapton PCBs. Since kapton is flexible,
the design of all-in-one kapton boards with long flexible tails as
cables has allowed in addition to place connectors, having
unacceptable activities (a few mBq per piece for isotopes of the
$^{238}$U and $^{232}$Th chains), behind the inner copper shielding.
NTC thermistors acting as temperature sensors and LEDs used for
calibration are also placed on the kapton boards; units fulfilling
NEXT requirements have been also selected. Solder paste of
acceptable~radiopurity has been found and will be used to fix SiPMs,
LEDs, and NTC sensors on kapton DBs.

The precise construction of NEXT-100 background model is underway
\cite{ichep}, based on Geant4 simulation. Using the activity levels
presented here, the estimated contribution of the tracking plane to
the background level in the region of interest for the neutrinoless
double beta decay of $^{136}$Xe has been preliminarily analyzed. As
it can be concluded from table~\ref{comcont}, the upper limits or
the quantified activity of $^{208}$Tl and $^{214}$Bi for the
selected components for the tracking readout give a rate below
$4\times10^{-5}$ counts keV$^{-1}$ kg$^{-1}$ y$^{-1}$, which is only
5\% of the required background level. A fruitful collaboration with
SensL company has allowed to improve the sensitivity of the
screening of SiPMs and then reduce considerably the upper limits
derived for the activity levels and therefore its impact on the
background model. The analysis of other components to be placed
behind the inner copper shield or even the vessel like connectors or
feedthroughs is also foreseen, although with lower priority.

\acknowledgments We deeply acknowledge John Murphy and Carl Jackson
from SensL Technologies Ltd for their efficient collaboration in the
analysis of SiPMs. We very much thank also Vicenzo Mancini from
SOMACIS company for the care in the development of radiopure kapton
PCBs. Special thanks are due to LSC directorate and staff for their
strong support for performing the measurements at the LSC
Radiopurity Service. The NEXT Collaboration acknowledges funding
support from the following agencies and institutions: the European
Research Council under the Advanced Grant 339787-NEXT and the T-REX
Starting Grant ref. ERC-2009-StG-240054 of the IDEAS program of the
7th EU Framework Program; the Spanish Ministerio de Economía y
Competitividad under grants CONSOLIDER-Ingenio 2010 CSD2008-0037
(CUP), FPA2009-13697-C04-04, and FIS2012-37947-C04; the Director,
Office of Science, Office of Basic Energy Sciences of the US DoE
under Contract no. DE-AC02-05CH11231; and the Portuguese FCT and
FEDER through the program COMPETE, Projects PTDC/FIS/103860/2008 and
PTDC/FIS/112272/2009.

\end{document}